\documentstyle[psfig]{mn}

\def\etal{{\it et al.\ }}
\def\eg{{\it e.g.\ }}

\def\spose#1{\hbox to 0pt{#1\hss}}
\def\approxlt{\mathrel{\spose{\lower 3pt\hbox{$\sim$}}
        \raise 2.0pt\hbox{$<$}}}
\def\approxgt{\mathrel{\spose{\lower 3pt\hbox{$\sim$}}
        \raise 2.0pt\hbox{$>$}}}
\def\approxpropto{\mathrel{\spose{\lower 3pt\hbox{$\sim$}}
        \raise 2.0pt\hbox{$\propto$}}}
\mathchardef\twiddle="2218

\def\multleft#1{\hbox to size{\vbox {\halign {\lft{##}\cr #1}}\hfill}\par}
\def\multright#1{\hbox to size{\vbox {\halign {\rt{##}\cr #1}}\hfill}\par}

\def\today{\ifcase\month\or January\or February\or March\or April\or May\or
      June\or July\or August\or September\or October\or November\or December\fi
      \space\number\day, \number\year}
\def\<{\thinspace}

\def\erg{{\rm\thinspace erg}}

\def\km{{\rm\thinspace km}}

\def\Mpc{{\rm\thinspace Mpc}}

\def\s{{\rm\thinspace s}}

\def\ergps{\hbox{$\erg\s^{-1}\,$}}

\def\kmps{\hbox{$\km\s^{-1}\,$}}

\def\kmpspMpc{\hbox{$\kmps\Mpc^{-1}$}}

\title[A preference for a non-zero neutrino mass from cosmological data]
{A preference for a non-zero neutrino mass from cosmological data}
\author[S.W. Allen et al.]
{\parbox[]{6.in} {S.W. Allen$^1$, R.W. Schmidt$^2$ and S.L. Bridle$^1$ \\
\footnotesize
1. Institute of Astronomy, Madingley Road, Cambridge CB3 0HA \\
2. Institut f\"ur Physik, Universit\"at Potsdam, Am Neuen
Palais 10, 14469 Potsdam, Germany \\
 }}
\begin{document}
\maketitle
\begin{abstract}
We present results from the analysis of cosmic microwave background
(CMB), large scale structure (galaxy redshift survey) and X-ray galaxy
cluster (baryon fraction and X-ray luminosity function) data, assuming
a geometrically flat cosmological model and allowing for tensor
components and a non-negligible neutrino mass.  From a combined
analysis of all data, assuming three degenerate neutrinos species, we
measure a contribution of neutrinos to the energy density of the
universe, $\Omega_{\nu}h^2=0.0059^{+0.0033}_{-0.0027}$ (68 per cent
confidence limits), with  zero falling on the 99 per cent confidence
limit. This corresponds to $\sim 4$ per cent of the total mass density
of the Universe and implies a species-summed neutrino mass $\sum_i m_i
=0.56^{+0.30}_{-0.26}$\,eV,  or $m_{\nu} \sim$ 0.2\,eV per neutrino.
We examine possible sources of systematic uncertainty in the
results. Combining the CMB, large scale structure and cluster baryon
fraction data, we measure an amplitude of mass fluctuations on
$8\,h^{-1}$Mpc scales of $\sigma_8=0.74^{+0.12}_{-0.07}$, which is
consistent with measurements based on the X-ray luminosity function
and other studies of the number density and evolution of galaxy
clusters. This value is lower than that obtained when fixing a
negligible neutrino mass ($\sigma_8=0.86^{+0.08}_{-0.07}$). The
combination of CMB, large scale structure and cluster baryon fraction
data also leads to remarkably tight constraints on the Hubble
constant, $H_0=68.4^{+2.0}_{-1.4}$\kmpspMpc, mean matter density,
$\Omega_{\rm m} =0.31\pm0.02$ and physical baryon density,
$\Omega_{\rm b}h^2=0.024\pm0.001$, of the Universe.
\end{abstract}


\section{Introduction}

The first results from the Wilkinson Microwave Anisotropy Probe (WMAP;
Bennett \etal 2003; Spergel \etal 2003) have confirmed, at high
statistical significance, what has become known as the `standard'
cosmological model. We appear to live in an approximately flat, vacuum
energy dominated universe, with a mean matter density $\Omega_{\rm m}
\sim 0.3$ and Hubble constant $H_0 \sim 70$\kmpspMpc, that was seeded
by predominantly adiabatic, Gaussian primordial fluctuations.

The initial papers from the WMAP team concentrated on results
extracted from the WMAP data alone, and on results obtained by 
combining the WMAP and other cosmic microwave background (CMB) data
with information from the 2dF Galaxy Redshift Survey (2dFGRS) and
observations of the Lyman-$\alpha$ forest. In this paper, we present
results based on similar CMB and 2dFGRS data,  but drop the
Lyman-$\alpha$ forest constraints.  Instead, we include constraints
provided by the baryonic mass fraction in X-ray luminous, dynamically
relaxed galaxy clusters studied with the Chandra Observatory, and
constraints obtained from the local  X-ray luminosity function (XLF)
of the most X-ray luminous galaxy clusters (Allen \etal 2002, 2003).

Neutrinos are fundamental particles of the Standard Model of particle
physics, and their masses have significant implications.  Experiments
to observe neutrino oscillations have shown there to be small, but
non-zero, mass squared differences between three neutrino mass
eigenstates (\eg Fukuda \etal 1998;  Ahmad \etal 2002; Eguchi \etal
2003).  If the lowest eigenstate has zero mass then these differences
imply a mass of $\sim 0.05$ eV for at least one of the remaining two
mass eigenstates.  Due to the effect of the neutrino mass on the
growth of structure in the Universe, cosmological observations are
sensitive to the neutrino mass density.  So far, only upper limits
have been obtained from cosmological observations.  The WMAP team
(Spergel et al. 2003) found $m_{\nu}<0.23$ eV per neutrino (95 per
cent confidence) from combining CMB, 2dFGRS and Lyman-$\alpha$ forest
data, assuming three neutrino species of degenerate mass (see Elgaroy
\& Lahav 2003 for a more detailed discussion of the assumptions).

Our primary motivation for this work was to compare and combine the
constraints from the new CMB and X-ray cluster data, with particular
attention being paid to the linearly-evolved amplitude of mass
fluctuations. Whereas the CMB data probe large to intermediate length
scales in the  early universe ($z\sim1100$), the XLF constraints are
much more local ($z\approxlt0.3$) and probe intermediate length
scales.  These different properties also make the combination of CMB
and XLF data well suited to examining  the contribution of neutrinos
to the mass density of the  universe.

The amplitude of spatial fluctuations in the mass density is a strong
function of both scale and epoch. Conventionally, the amplitude of
fluctuations is quoted on an intermediate length scale at the present
day (z=0): $\sigma_8$ is the root-mean-square matter fluctuation
within $8\,h^{-1}$\,Mpc  spheres, calculated using linear theory. As
discussed  by e.g. Efstathiou \& Bond (1999), there are a large number
of different  (degenerate) combinations of parameters that can give
almost identical  CMB power spectra, most importantly (assuming a flat
universe) the  Hubble constant, $H_0 = 100\,h$\kmpspMpc,  the
neutrino mass density, $\Omega_{\nu}$, and the mean matter density in
units of the critical density, $\Omega_{\rm m}$.  Current CMB data
alone cannot, therefore, provide strong  constraints on $\sigma_8$
(Spergel \etal 2003, Bridle \etal 2003).

Although the normalization of the CMB power spectrum  on large scales
is insensitive to the neutrino mass,  allowing for the presence of
massive neutrinos will lower the value of $\sigma_8$ inferred  from the
CMB data. This is because massive neutrinos  affect the growth of
fluctuations on small scales at late  times ($\sigma_8$ being
calculated at the present day).  Conversely, whereas the number
density of galaxy clusters in the  local universe is sensitive to the
neutrino mass, the value of  $\sigma_8$ inferred from the XLF
observations is only mildly  affected, since the length scale probed
by clusters  is very similar to $8\,h^{-1}$ Mpc.

Given an accurate measurement of $\Omega_{\rm m}$, the  cluster XLF
data can yield a precise determination of $\sigma_8$.  $\Omega_{\rm
m}$ can be obtained  from measurements of the baryonic gas mass
fraction, $f_{\rm
gas}$, in large,  dynamically relaxed clusters of galaxies, together with  results on the Hubble constant and mean baryon
density of  the Universe from the ratio of first to second peak
heights in the CMB temperature  angular power spectrum. Thus, the
combination of CMB, $f_{\rm gas}$ and  cluster XLF data can together
yield an estimate of the neutrino mass.

Our approach in constraining the neutrino mass is similar to that
described by Fukugita, Liu \& Sugiyama (2000). A number of earlier
papers have also placed limits on the neutrino mass using the observed
slope of the galaxy (and by assumption matter) power spectrum
obtained from the 2dFGRS, which is sensitive  to a combination of
cosmological parameters  (see Elgaroy \etal 2002, Elgaroy \& Lahav
2003, Hannestad 2003 and references therein). We show that including
the 2dFGRS constraint is helpful in constraining the neutrino  mass,
but not essential to our analysis.

\section{Method}

\subsection{Analysis of the CMB and large scale structure data}

Our analysis of CMB observations uses the WMAP temperature (TT) data
for multipoles $l<900$ (Hinshaw \etal 2003) and
temperature-polarization (TE) data for $l<450$ (Kogut \etal 2003).  To
extend the analysis to higher multipoles (smaller scales), we also
include data from the Cosmic Background Imager (CBI; Pearson \etal
2003) and Arcminute Cosmology Bolometer Array Receiver (ACBAR; Kuo
\etal 2003) for $l > 800$. The comparison of model angular power
spectra with the WMAP data employs the likelihood calculation routines
released  by the WMAP team (Verde \etal 2003).

For part of the analysis, we also include large scale structure data
from the 2dFGRS.  We use the power spectrum obtained from the first
147,000 redshifts of the survey, covering scales $0.02<k/(h\,{\rm
Mpc}^{-1})<0.15$, where non-linear effects are thought to be
negligible (Percival \etal 2002). Over this range, we assume that  the
2dFGRS power spectrum is directly proportional to the matter power
spectrum at $z=0$, marginalizing over the constant of proportionality
using a wide prior.

Our analysis of the CMB and 2dFGRS data uses the CosmoMC
code\footnote{{http://cosmologist.info/cosmomc/}}. This in turn
uses CAMB (Lewis, Challinor \& Lasenby 2000), which is based on
CMBFAST (Seljak \& Zaldarriaga 1996), to generate the CMB and matter
power spectrum transfer functions, and a Metropolis-Hastings Markov
Chain Monte Carlo (MCMC) algorithm to explore parameter space. We used
the covariance matrix of the parameters to improve sampling efficiency
(see Lewis \& Bridle 2002 for  more details). 

We have fitted the data using a cosmological model with nine free
parameters: the physical dark (cold+hot) matter and baryon densities in
units of the critical density, $\Omega_{\rm dm}h^2$ and $\Omega_{\rm
b}h^2$, the Hubble constant $H_0$,  the neutrino mass fraction
$f_\nu=\Omega_{\nu}/\Omega_{\rm dm}$,  the recombination redshift
$z_{\rm rec}$ (at which the reionization fraction is a half, assuming
instantaneous reionization), the amplitude of the scalar power
spectrum $A_{\rm S}$, the scalar and tensor spectral indexes $n_{\rm
S}$ and $n_{\rm T}$,  and the ratio of the normalization of tensor and
scalar components $R=A_{\rm T}/A_{\rm S}$, where $A_{\rm S}$ is
evaluated at $k=0.05 {\rm Mpc}^{-1}$ and $A_{\rm T}$ is evaluated at
$k=0.002 {\rm Mpc}^{-1}$.  We assume three neutrinos of degenerate
(equal) mass throughout.  Each parameter had a wide uniform prior
applied. We assume a flat geometry with a cosmological constant
($\Omega_{\rm m}+\Omega_{\Lambda}$=1).

The analysis was carried out on the Cambridge X-ray group Linux
cluster. For our main analyses of the CMB and CMB+2dFGRS data, using the
nine-parameter model and allowing for massive neutrinos and tensors,
we accumulated at total of $10^6$ correlated  samples in 10 separate
chains. In order to investigate the effects of neutrinos and tensors
in detail, we also carried out two additional runs in which the
neutrino density and tensor amplitude ratio were separately fixed to
zero ($f_\nu=0$ and $R=0$, respectively). For these runs, we
accumulated $6\times 10^5$ samples in 6 separate chains.  We satisfied
ourselves that the chains had converged by ensuring that consistent
final results were obtained from numerous small subsets of the
chains. In all cases, we allowed a conservative burn-in period of
$10^4$ samples.

\subsection{Importance sampling using the X-ray cluster constraints}

\begin{figure}
\vspace{0.5cm} \hbox{
\hspace{-0.1cm}\psfig{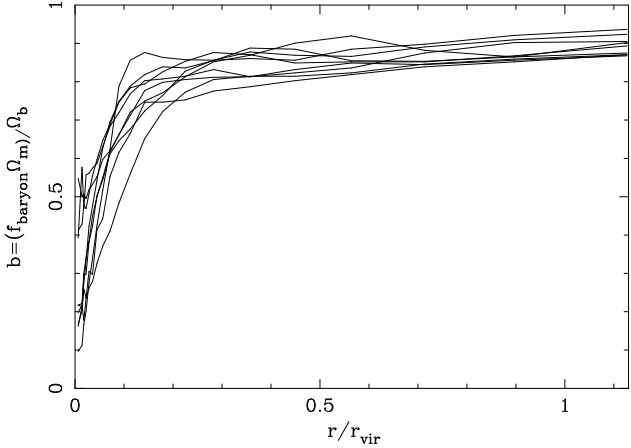}}
\caption{The enclosed baryon fraction relative to the universal value
as a function of radius, in units of the virial radius $r_{\rm vir}$,
from the  simulations of Eke \etal (1998). The simulated clusters have
similar masses and temperatures to the clusters used here. The results
for the most dynamically active system  in the simulations have been
excluded. Beyond a radius  $r > 0.2 r_{\rm vir}$, the simulated
clusters exhibit consistent,  relatively flat $b$ profiles.  At
$r=0.25 r_{\rm vir}$, a radius comparable  to the measurement radius
for the Chandra observations, the simulations  give
$b=0.824\pm0.033$.}\label{fig:fgas}
\end{figure}

In order to obtain improved results on cosmological parameters we
have importance sampled the CMB and CMB+2dFGRS MCMC results, applying 
further, independent constraints based on X-ray studies of galaxy clusters.
Firstly, we applied constraints from the observed X-ray gas
mass fraction, $f_{\rm gas}$, in dynamically relaxed  clusters. These
data provide one of the most simple and  robust cosmological probes
currently available, constraining a combination of $\Omega_{\rm
m}$, $H_0$ and $\Omega_{\rm b}$.  We have used the
data and methods of Allen, Schmidt \& Fabian (2002) and Allen \etal
(2003).  The data are drawn from Chandra observations of X-ray
luminous,  dynamically relaxed clusters spanning the redshift range
$0.08<z<0.47$. Consistent measurements of the mass profiles in the
clusters are available from X-ray and  gravitational lensing methods
in most cases, keeping systematic uncertainties to a minimum.

Each of the MCMC samples obtained from the CMB and 
CMB+2dFGRS analyses
provides
values for $\Omega_{\rm b}$,  $\Omega_{\rm m}$ and $h$. Using these
values, we fit the observed $f_{\rm gas}(z)$ data with the model
function (Allen \etal 2002)

\begin{eqnarray}
f_{\rm gas}^{\rm mod}(z) = \frac{ b\, \Omega_{\rm b}} {\left(1+0.19
\sqrt{h}\right) \Omega_{\rm m}} \left[ \frac{h}{0.5} \, \frac{D_{\rm
A}^{\Omega_{\rm m}=1, \Omega_{\rm  \Lambda}=0 }(z)}{D_{\rm
A}^{\Omega_{\rm m},\,\Omega_{\Lambda}}(z)} \right]^{1.5},
\end{eqnarray}

\noindent where $\Omega_{\rm m}=1.0$, $\Omega_{\Lambda}=0$ and 
$h=0.5$ define the reference cosmology used for the 
$f_{\rm gas}(z)$ measurements. The X-ray gas 
dominates the baryonic mass content of the clusters and
the factor $(1+0.19 \sqrt{h})$ converts the X-ray gas  mass into the
total baryonic mass (Fukugita, Hogan \&  Peebles 1998). The quantity
$b$ is a bias factor, motivated by gasdynamical simulations which
suggest that the baryon fraction in clusters is slightly lower than
for the universe as a whole (\eg Cen \& Ostriker 1994; Eke, Navarro \&
Frenk 1998; Bialek \etal  2001). We use the results of Eke \etal
(1998) from simulations of 10 X-ray luminous,  massive clusters to
constrain $b$.  Excluding the data for the most dynamically active
cluster in the study of Eke \etal (1998; recall that the $f_{\rm gas}$
data are drawn from Chandra observations of  dynamically relaxed
systems), the simulated clusters show consistent, relatively flat
baryonic mass fraction profiles for radii $r > 0.2 r_{\rm vir}$ 
(Fig~\ref{fig:fgas}).  At $r=0.25 r_{\rm vir}$, a radius comparable to
the measurement radius  for the Chandra observations, the simulations
of  Eke \etal (1998) give $b=0.824\pm0.033$.

Employing a Gaussian prior on the bias factor and fitting the model
described by Eq. 1, we obtain a $\chi^2$ value for each of the
MCMC samples.  The weight of the MCMC samples is then multiplied by
$e^{-\chi^2/2}$. Note that we assume that the effect of  massive
neutrinos on the $f_{\rm gas}$ constraints is negligible,  which would
be true if the neutrino to dark matter ratio within the galaxy
clusters were typical of the universe as a whole. However, even in the
most extreme case in which neutrinos completely avoid galaxy clusters,
$\Omega_{\rm m}$ in the pre-factor of Eq.~1 need only be replaced by 
$\Omega_{\rm m} - \Omega_{\nu}$.  Since we find
$\Omega_{\nu}/\Omega_{\rm m}\sim 0.04$ from this work, even this
extreme case would only increase $\Omega_{\rm m}$ by $\sim 0.01$ and
have negligible effect on our conclusions regarding the neutrino mass.

Secondly, we have also importance sampled the MCMC results
applying additional constraints  from the observed, local X-ray
luminosity function (XLF)  of galaxy clusters. For this analysis, we
use the results of Allen \etal (2003; see Fig. 3 of that paper), which
are  based on the XLF for the most X-ray luminous ($L_{\rm
X,0.1-2.4}>10^{45}$\ergps) galaxy clusters  within redshift $z<0.3$
observed in the ROSAT All-Sky Survey (Ebeling \etal 2000;  B\"ohringer
\etal 2002), and a mass-luminosity relation  determined from Chandra
X-ray and weak lensing observations.  The XLF results constrain the
combination of  $\Omega_{\rm m}$ and $\sigma_8$. To importance sample,
we multiply the weight of  each of the MCMC samples by the probability
of that $\sigma_8$, $\Omega_{\rm m}$ parameter pair from the
analysis of Allen \etal (2003).

\begin{figure*}
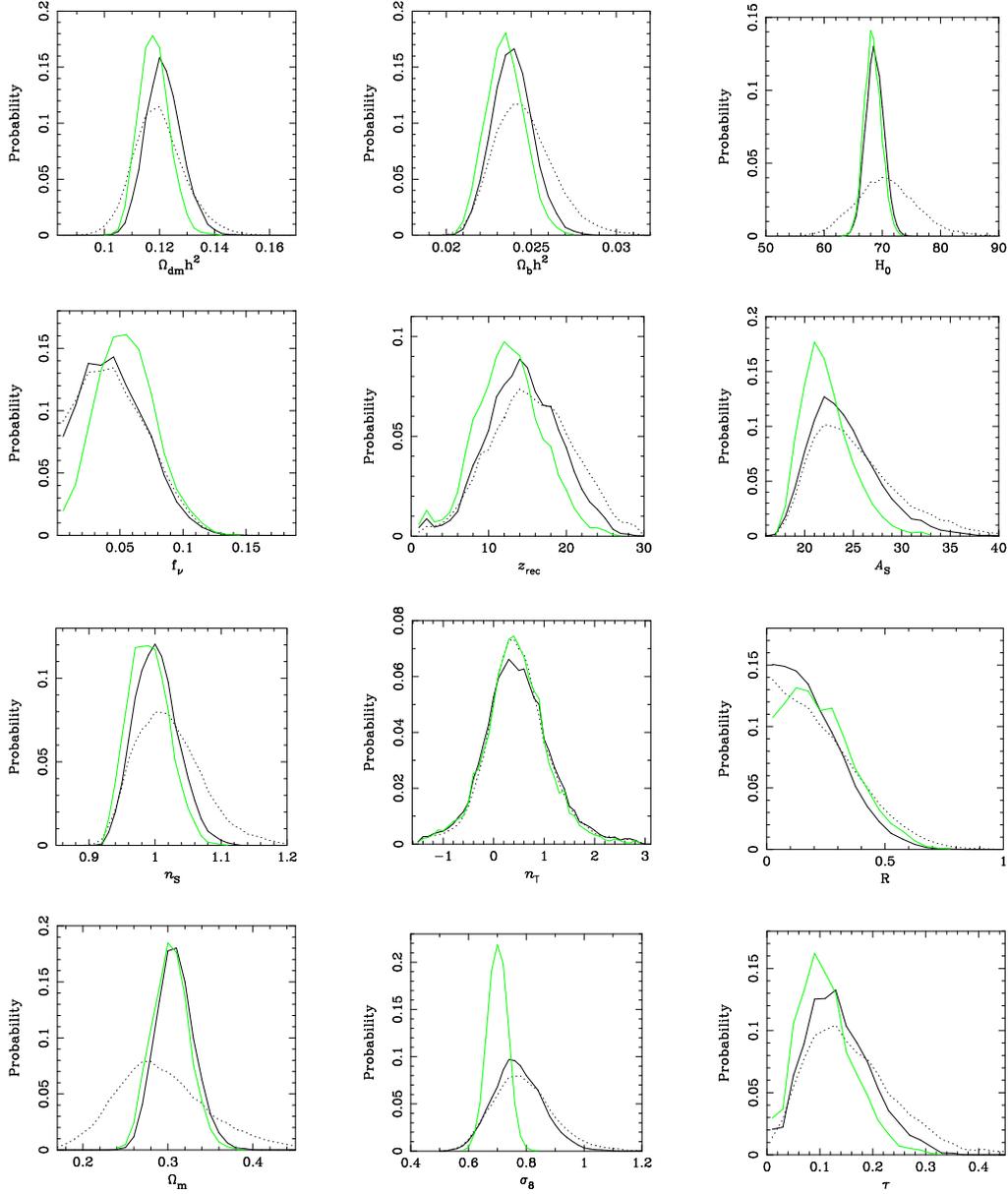

\vspace{0.2cm} 
\hbox{
\hspace{2.0cm}\psfig{figure=fig2a.ps,width=0.22
\textwidth,angle=270}
\hspace{0.8cm}\psfig{figure=fig2b.ps,width=0.22
\textwidth,angle=270}
\hspace{0.8cm}\psfig{figure=fig2c.ps,width=0.22
\textwidth,angle=270} } 
\vspace{0.4cm} 
\hbox{
\hspace{2.0cm}\psfig{figure=fig2d.ps,width=0.22
\textwidth,angle=270}
\hspace{0.8cm}\psfig{figure=fig2e.ps,width=0.22
\textwidth,angle=270}
\hspace{0.8cm}\psfig{figure=fig2f.ps,width=0.22
\textwidth,angle=270} } 
\vspace{0.4cm} 
\hbox{
\hspace{2.0cm}\psfig{figure=fig2g.ps,width=0.22
\textwidth,angle=270}
\hspace{0.8cm}\psfig{figure=fig2h.ps,width=0.22
\textwidth,angle=270}
\hspace{0.8cm}\psfig{figure=fig2i.ps,width=0.22
\textwidth,angle=270} } 
\vspace{0.4cm} 
\hbox{
\hspace{2.0cm}\psfig{figure=fig2j.ps,width=0.22
\textwidth,angle=270}
\hspace{0.8cm}\psfig{figure=fig2k.ps,width=0.22
\textwidth,angle=270}
\hspace{0.8cm}\psfig{figure=fig2l.ps,width=0.22
\textwidth,angle=270} }
\vspace{0.4cm} 
\caption{Marginalized probability distributions for the cosmological 
parameters from the analysis of the CMB+2dF (dotted) CMB+2dF+$f_{\rm
gas}$ (dark solid) and CMB+2dF+$f_{\rm
gas}$+XLF (grey solid) data using the nine-parameter model, allowing 
for the presence of tensor components and massive neutrinos.}\label{fig:marginal}
\end{figure*}

\section{Results} 

The marginalized results on cosmological  parameters obtained from the
analysis of the CMB+\linebreak[0]2dF,
CMB+\linebreak[0]2dF+\linebreak[0]$f_{\rm gas}$,
CMB+\linebreak[0]2dF+\linebreak[0]XLF,  
CMB+\linebreak[0]$f_{\rm gas}$+\linebreak[0]XLF,  
and CMB+\linebreak[0]2dF+\linebreak[0]$f_{\rm gas}$+\linebreak[0]XLF data,
using the nine parameter model with the neutrino mass fraction and
tensor amplitude ratio included as free parameters, are summarized
in Table~\ref{table:marginal1}. This table also lists the values of
other quantities that can be derived from the  parameter values: the
matter density in  units of the critical density, $\Omega_{\rm m}$ ($\Omega_{\rm m}=\Omega_{\rm dm}+\Omega_{\rm b}$),
the linearly-evolved  amplitude of mass fluctuations on 8$h^{-1}$Mpc
scales, $\sigma_8$, the physical neutrino density in units of the
critical density, $\Omega_{\nu}h^2$, and the optical depth to
reionization, $\tau$. 

The marginalized, posterior probability
distributions for the parameters are also shown in
Fig.~\ref{fig:marginal}. The results obtained from the additional runs
with the neutrino mass fraction, and then the tensor amplitude ratio,
fixed to zero are, in general, consistent with those listed in
Table~\ref{table:marginal1}. The most notable exception to this is the
result on $\sigma_8$ in the  case of a negligible neutrino mass, 
which is discussed  below. We note, however, that the run without
tensors gives peak-probability values for $\Omega_{\rm b}h^2$, $H_0$,
$f_{\nu}$, $n_{\rm S}$ and $\Omega_{\nu}h^2$ that are 
slightly lower (and a result on $\sigma_8$ that is slightly higher) 
than the tabulated values, although consistent within 68 per cent 
uncertainties. We also note that the limit on the ratio of the scalar to tensor
amplitudes is constrained more tightly in the $f_{\nu}=0$  case
($R<0.3$ at 95 per cent confidence).

\begin{table*}
\begin{center}
\caption{The marginalized results on cosmological parameters obtained 
using the nine-parameter model, allowing for tensor components and massive 
neutrinos. The quoted values correspond  to the peaks
of the marginalized probability distributions.  Error bars are 68.3
per cent confidence limits.  Where only a limit is quoted, this
corresponds to the 95.4 per cent confidence
level.}\label{table:marginal1}
\begin{tabular}{c c c c c c c }
\multicolumn{1}{c}{} &
\multicolumn{1}{c}{~~} &
\multicolumn{5}{c}{} \\

                        & ~ & CMB+2dF                      &  CMB+2dF+$f_{\rm gas}$       &  CMB+2dF+XLF                 &  CMB+$f_{\rm gas}$+XLF                 &  CMB+2dF+$f_{\rm gas}$+XLF    \\
\hline                                                                                                                                                                                  
$\Omega_{\rm dm}h^2$    & ~ & $0.120^{+0.008}_{-0.010}$    &  $0.121^{+0.007}_{-0.007}$   &  $0.115^{+0.008}_{-0.008}$   &  $0.118^{+0.006}_{-0.006}$   &  $0.118^{+0.005}_{-0.006}$    \\
$\Omega_{\rm b}h^2$     & ~ & $0.0240^{+0.0021}_{-0.0016}$ &  $0.0238^{+0.0012}_{-0.0012}$&  $0.0236^{+0.0018}_{-0.0014}$&  $0.0236^{+0.0012}_{-0.0013}$&  $0.0236^{+0.0010}_{-0.0013}$ \\ 
$H_0$                   & ~ & $70^{+6}_{-5}$               &  $68.4^{+2.0}_{-1.4}$        &  $71.5^{+3.5}_{-6.0}$        &  $67.6^{+2.2}_{-1.1}$        &  $68.4^{+1.5}_{-1.6}$         \\
$f_{\nu}$               & ~ & $0.042^{+0.023}_{-0.036}$    &  $0.042^{+0.024}_{-0.029}$   &  $0.046^{+0.031}_{-0.020}$   &  $0.042^{+0.053}_{-0.010}$   &  $0.051^{+0.026}_{-0.025}$    \\
$z_{\rm rec}$           & ~ & $14^{+7}_{-4}$               &  $14^{+5}_{-4}$              &  $13^{+5}_{-5}$              &  $10^{+11}_{-3}$              &  $12^{+4}_{-4.5}$             \\
$A_{\rm S}$             & ~ & $22^{+6}_{-3}$               &  $22.0^{+4.5}_{-2.5}$        &  $21.5^{+3.5}_{-2.5}$        &  $20.0^{+6.0}_{-1.5}$        &  $21.5^{+2.5}_{-2.5}$         \\
$n_{\rm S}$             & ~ & $1.00^{+0.06}_{-0.04}$       &  $1.00^{+0.03}_{-0.04}$      &  $1.00^{+0.05}_{-0.05}$      &  $1.00^{+0.03}_{-0.05}$      &  $0.98^{+0.04}_{-0.03}$       \\ 
$n_{\rm T}$             & ~ & $0.4\pm0.6$                  &  $0.3^{+0.7}_{-0.5}$         &  $0.4^{+0.5}_{-0.5}$         &  $0.2^{+0.8}_{-0.4}$         &  $0.4^{+0.6}_{-0.6}$          \\
$R$                     & ~ & $<0.6$                       &  $<0.5$                      &  $<0.6$                      &  $<0.5$                      &  $<0.5$                       \\
&&&&&&\\                                                                                                                                                                                    
$\Omega_{\rm m}$        & ~ & $0.28^{+0.06}_{-0.05}$       &  $0.308^{+0.020}_{-0.024}$   &  $0.27^{+0.06}_{-0.03}$      &  $0.296^{+0.036}_{-0.020}$      &  $0.301^{+0.024}_{-0.024}$     \\ 
$\sigma_8$              & ~ & $0.78^{+0.10}_{-0.11}$       &  $0.74^{+0.12}_{-0.07}$      &  $0.72^{+0.04}_{-0.06}$      &  $0.69^{+0.04}_{-0.05}$      &  $0.70^{+0.04}_{-0.04}$        \\
$\Omega_{\nu}h^2$       & ~ & $0.0040^{+0.0035}_{-0.0035}$ &  $0.0045^{+0.0036}_{-0.0031}$&  $0.0055^{+0.0034}_{-0.0029}$&  $0.0055^{+0.0070}_{-0.0017}$&  $0.0059^{+0.0033}_{-0.0027}$  \\ 
$\tau$                  & ~ & $0.13^{+0.09}_{-0.08}$       &  $0.13^{+0.06}_{-0.07}$      &  $0.09^{+0.08}_{-0.05}$      &  $0.07^{+0.15}_{-0.03}$      &  $0.09^{+0.06}_{-0.05}$        \\
 \hline
\end{tabular}
\end{center}
\end{table*}

\subsection{The cosmic neutrino density and amplitude of mass fluctuations}\label{section:neutrinos}

\begin{figure}
\vspace{0.5cm} \hbox{
\hspace{0.2cm}\psfig{figure=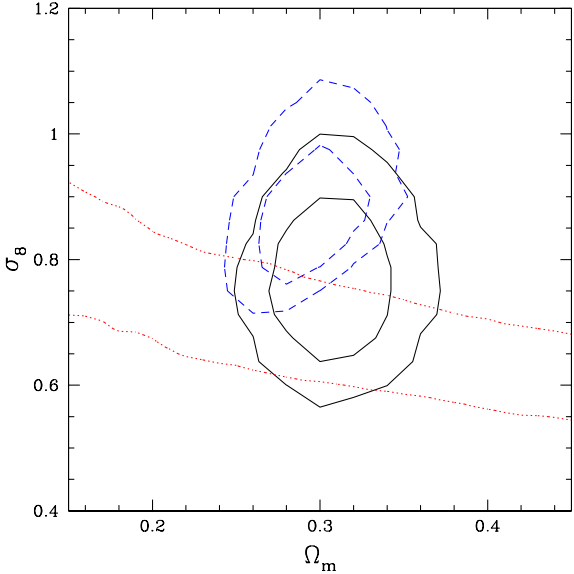,width=.45\textwidth,angle=0}}
\caption{Joint 68.3 and 95.4 per cent confidence limits 
on $\sigma_8$ and $\Omega_{\rm m}$ from the analysis of the
CMB+2dF+$f_{\rm gas}$ data, allowing (solid curve) and excluding
(dashed curve) a contribution to the energy density of the universe
from massive neutrinos. The dotted curve shows the 95.4 per cent
confidence limit from the analysis of the local cluster XLF from
Allen \etal 2003}\label{fig:nucont}
\end{figure}

Two of the more interesting results to come from the analysis are 
the effects of allowing for massive neutrinos on the present day, 
linearly-evolved amplitude of mass fluctuations, $\sigma_8$, and 
the implied contribution of such neutrinos to the energy 
density of the  universe, $\Omega_{\nu}h^2$.

Fig.~\ref{fig:nucont} shows the joint 68.3 and 95.4 per cent
confidence limits on $\Omega_{\rm m}$ and $\sigma_8$ obtained from the
CMB+2dF+$f_{\rm gas}$ data. The independent constraints from the
cluster XLF are also shown overlaid. In the absence of massive
neutrinos ($f_{\nu}=0$), we obtain a marginalized result on  the
amplitude of mass fluctuations from the CMB+2dF+$f_{\rm gas}$ data of
$\sigma_8=0.86^{+0.08}_{-0.07}$ (dashed curve in
Fig.~\ref{fig:nucont}), which is only marginally consistent with the
cluster XLF constraints. When the neutrino density is included as a
free parameter, however, the agreement with the XLF results is
improved significantly, with a marginalized result on $\sigma_8$ from
the CMB+2dF+$f_{\rm gas}$ data of $\sigma_8=0.74^{+0.12}_{-0.07}$
(solid curve in Fig.~\ref{fig:nucont}. A similar degeneracy between
$\sigma_8$ and $f_{\nu}$ is visible in the analysis of pre-WMAP CMB
(and other) data by Lewis \& Bridle 2003).  We note that the effects of
neutrinos on the cluster XLF analysis are negligible, reducing the
measured value of $\sigma_8$ by only $\sim 1.5$ per cent from the
no-neutrino case (see Section~\ref{section:discussion}).

The marginalized results on $\sigma_8$ from the
CMB+\linebreak[0]2dF+\linebreak[0]$f_{\rm gas}$ data are also
consistent with those obtained from other recent studies of the number
density and evolution of galaxy clusters, which generally find
$\sigma_8 \sim 0.65-0.75$ for $\Omega_{\rm m} = 0.3$ (\eg Borgani
\etal 2001; Reiprich \& B\"ohringer 2002; Viana, Nichol \& Liddle
2002; Schuecker \etal 2003; Pierpaoli \etal 2003; Voevodkin \&
Vikhlinin 2003 and references therein) as well as some of the more
recent measurements  of cosmic shear on large angular scales (Brown
\etal 2003; Jarvis \etal  2003 and references therein). We caution,
however, that some  other cosmic shear  studies (\eg Van Waerbeke
\etal 2002; Refregier, Rhodes \& Groth 2002) have  inferred higher
values of $\sigma_8 \sim 0.9$, for $\Omega_{\rm m} \sim 0.3$.

\begin{figure*}
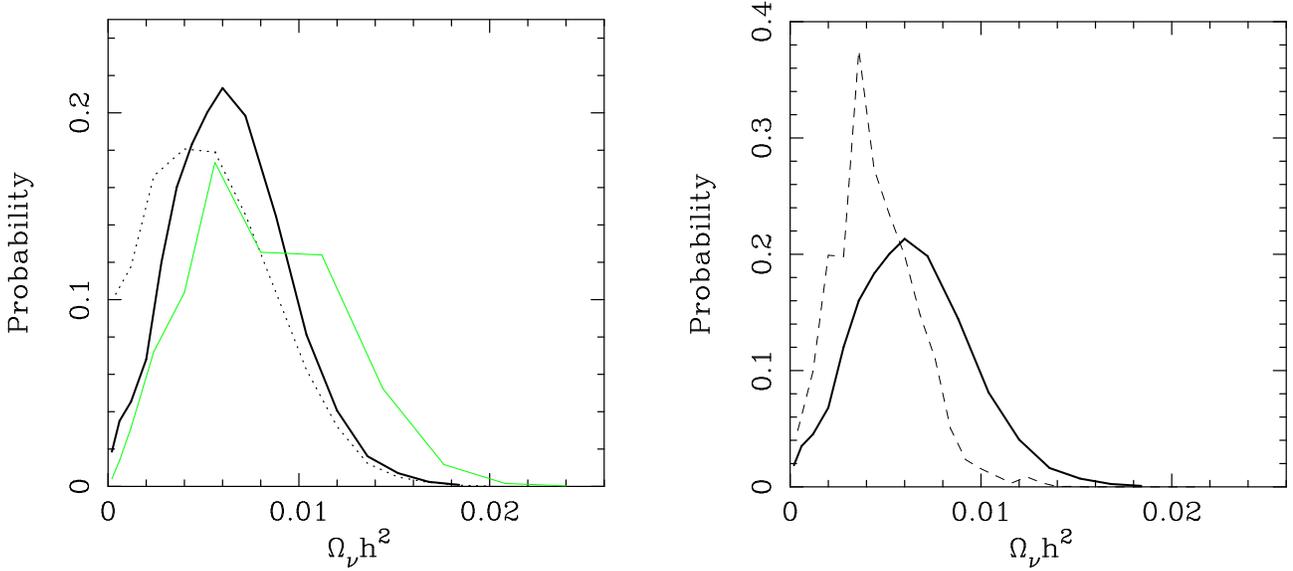

\vspace{0.5cm} \hbox{
\hspace{0.2cm}\psfig{figure=fig4a.ps,width=.45\textwidth,angle=270}
\hspace{1.0cm}\psfig{figure=fig4b.ps,width=.45\textwidth,angle=270}}
\caption{(a) Marginalized probability distribution for 
$\Omega_{\nu}h^2$ from the analysis of  the CMB+2dF+$f_{\rm gas}$
(dotted curve),  CMB+$f_{\rm gas}$+XLF data (grey curve) and
CMB+2dF+$f_{\rm gas}$+XLF (dark, solid curve) data. In all cases we
have  allowed for the presence of tensor components (b) The dashed
curve shows the results from the CMB+2dF+$f_{\rm gas}$+XLF data from
the  analysis without tensor components. The dark, solid curve is  the
same as in (a).}\label{fig:omeganuh2}
\end{figure*}

Fig.~\ref{fig:omeganuh2}(a) shows the constraints on $\Omega_{\nu}h^2$
obtained from  various combinations of the data, as described in
Table~\ref{table:marginal1}.  Using the CMB+2dF+$f_{\rm gas}$ data, we
measure a neutrino density
$\Omega_{\nu}h^2=0.0045^{+0.0036}_{-0.0031}$, with a 95.4 per cent
confidence upper limit of $\Omega_{\nu} h^2<0.012$ (or $f_{\nu}<0.1$;
the same upper limit is also obtained using only the CMB+2dF data.)
For the full
CMB+2dF+$f_{\rm gas}$+XLF data set,  we obtain
$\Omega_{\nu}h^2=0.0059^{+0.0033}_{-0.0027}$, with the region between 
$\Omega_{\nu}h^2=0$ and the corresponding equi-probability upper value
for $\Omega_{\nu}h^2$ containing 99 per cent of the probability. Finally,
for the CMB+$f_{\rm gas}$+XLF data, we find
$\Omega_{\nu}h^2=0.0055^{+0.0070}_{-0.0017}$, with $\Omega_{\nu}h^2>0$
at more than 99 per cent confidence. We thus see that including  the
2dFGRS constraint is helpful in the analysis, but not essential  to
our detection of a non-zero neutrino mass. Our most probable result on
the neutrino density from the CMB+$f_{\rm gas}$+XLF  and
CMB+2dF+$f_{\rm gas}$+XLF data sets corresponds  to $\sim 1.3$ per
cent of the critical value required for closure.

Fig.~\ref{fig:omeganuh2}(b) shows how the results  on
$\Omega_{\nu}h^2$ are modified when we exclude the presence of tensor
components. In this case, the full  CMB+2dF+$f_{\rm gas}$+XLF data set
gives  $\Omega_{\nu}h^2=0.0034^{+0.0031}_{-0.0016}$. Although the most
probable value for $\Omega_{\nu}h^2$ is reduced when we exclude tensor
components, we still find $\Omega_{\nu}h^2>0$ at 93 per cent confidence.

The results on $\Omega_{\nu}h^2$ can be converted into a constraint on
the neutrino mass, summed over species, using the relation
$\Omega_{\nu} h^2 = {\sum_i m_i}/{94\,{\rm eV}}$.  For the  full
CMB+2dF+$f_{\rm gas}$+XLF data set, we find $\sum_i m_i  = 
0.56^{+0.30}_{-0.26}$eV.

Our result on the contribution of neutrinos to the energy density of
the universe differs slightly from that reported by the WMAP team,
$\Omega_{\nu} h^2<0.0076$ (95 per cent confidence limit; Spergel \etal
2003).  The primary reason for this difference is the inclusion of
Lyman-$\alpha$ forest data in the WMAP team's analysis (see
discussions in Elgaroy \& Lahav 2003 and Hannestad 2003) and the
inclusion of cluster XLF data in ours. Our results are,  however,
consistent with those of  2dFGRS team: $\sum_i m_i < 2.2$ eV  (Elgaroy
\etal 2002; Hannestad 2003) as well as those of Lewis \& Bridle (2003)
and  Slosar \etal (2003) from the analysis of pre-WMAP CMB, 2dFGRS and
other data using MCMC techniques.

\subsection{The Hubble constant and mean matter density}

Our results on the Hubble constant are in good agreement with, but
more tightly constrained than, that from the Hubble Key Project
($H_0=72\pm8$\kmpspMpc; Freedman \etal 2001). Including the neutrino
energy density as a free parameter, our analysis of the CMB+2dF data
gives $H_0=70^{+6}_{-5}$\kmpspMpc. The introduction of the Chandra
$f_{\rm gas}$ data improves this constraint significantly, to
$H_0=68.4^{+2.0}_{-1.4}$\kmpspMpc.  For the full CMB+2dF+$f_{\rm
gas}$+XLF data set, we find $H_0=68.4^{+1.5}_{-1.6}$\kmpspMpc.

\begin{figure}
\vspace{0.2cm}
\hbox{
\hspace{0.2cm}\psfig{figure=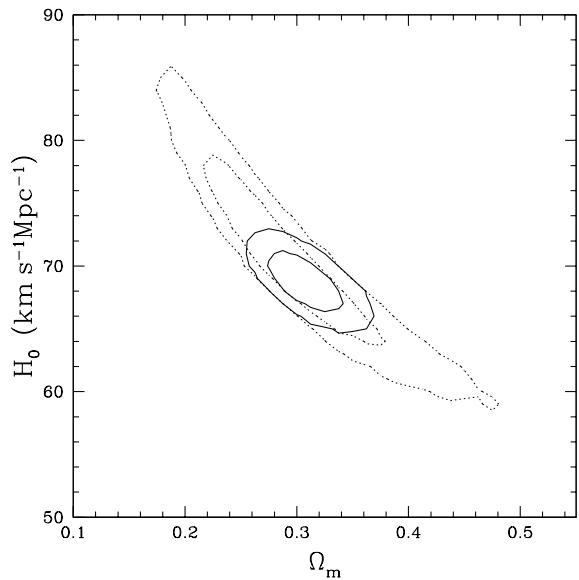,width=0.45 \textwidth,angle=0}
}
\caption{The joint 68.3 and 95.4 per cent confidence limits 
on $H_0$ and $\Omega_{\rm m}$  from the analysis of the CMB+2dF
(dotted curve) and  CMB+2dF+$f_{\rm gas}$ (solid curve) data. 
We have allowed for the presence of tensor 
components and massive neutrinos in the analysis.}\label{fig:omh0}
\end{figure}

The introduction of the $f_{\rm gas}$ data also leads to significant
improvements in our measurements of the mean matter and baryon
densities. As can be seen from Table~\ref{table:marginal1}, the
combination of CMB+2dF data gives a $\sim 20$ per cent uncertainty on
$\Omega_{\rm m}$ and an eight per  cent uncertainty on $\Omega_{\rm
b}h^2$.  The introduction of the $f_{\rm gas}$ data improves these
constraints to seven per cent on $\Omega_{\rm m}$ and five  per cent
on $\Omega_{\rm b}h^2$. The further inclusion of the XLF data makes
little difference to the results, providing us with final
answers of $\Omega_{\rm m}= 0.301 \pm 0.024$ and $\Omega_{\rm  b} h^2
= 0.0236 \pm 0.0012$.  The improvement in the constraints on  $H_0$
and $\Omega_{\rm m}$ obtained with the introduction of the Chandra
$f_{\rm gas}$ data can be clearly seen in Fig.~\ref{fig:omh0}.
 
We note that the tight constraints on $H_0$, $\Omega_{\rm m}$ and
$\Omega_{\rm b}h^2$ are primarily due to the combination of CMB and
$f_{\rm gas}$ data. From an analysis of these two data sets alone, we
find $H_0=66.4^{+2.4}_{-1.5}$\kmpspMpc, $\Omega_{\rm
m}=0.304^{+0.024}_{-0.027}$ and $\Omega_{\rm
b}h^2=0.0234^{+0.0015}_{-0.0018}$.

\section{Discussion}\label{section:discussion}

The main result of this paper is the tentative detection of a
non-negligible neutrino mass. Recent years have seen significant
progress in our understanding of neutrinos from particle physics. The
Large Electron Positron Collider (LEP) and Stanford Linear Accelerator
Centre Linear Electron-Positron Collider (SLC) showed there to be
three light, active neutrino species (electron, muon,  tau). Solar and
atmospheric neutrino experiments such as Super-Kamiokande, the Sudbury
Neutrino Observatory and Kamland (Fukuda \etal 1998;  Ahmad \etal
2002; Eguchi \etal 2003) have since provided strong evidence for
oscillations between these neutrino species, and determined two of the
mass-squared  differences to be ${\Delta m^2_{\rm sol}} \sim 7 \times
10^{-5}$ eV$^2$ and ${\Delta m^2_{\rm atm}} \sim 3\times 10^{-3}$
eV$^2$. Given the relatively small mass-squared differences, the
result on the species-summed neutrino mass presented here, $\sum_i
m_i= 0.56^{+0.30}_{-0.26}$\,eV (for the CMB+2dF+$f_{\rm gas}$+XLF
data)  implies an approximately degenerate neutrino mass. Writing
${\sum_i m_i} =  N m_{\nu}$ and setting $N=3$ (assuming an absence of
further, heavy sterile neutrino species), we obtain $m_\nu
=0.19^{+0.10}_{-0.09}$\,eV.  Note, however, that our results are also
marginally consistent with indications from the Liquid Scintillator
Neutrino Detector experiment (Aguilar \etal 2001) for the presence of
sterile neutrinos.

Our results are consistent with the upper limit on the electron
neutrino mass, $m_{\nu} < 2.2$ eV (95 per cent confidence limit), from
laboratory tritium beta decay end-point experiments (Weinheimer \etal
1999;  Bonn \etal 2002). We note that our results are also consistent
with the neutrino mass inferred from Z-burst models (Fargion, Mele \&
Salis 1999; Weiler 1999) for ultrahigh energy cosmic rays,
$m_{\nu}=0.26^{+0.20}_{-0.14}$\,{\rm eV}  (Fodor, Katz  \& Ringwald
2001; assuming an  extragalactic origin for the cosmic rays), some
indications from  neutrinoless double beta decay experiments,
$0.11\,{\rm eV} \leq m_{\nu} \leq 0.56$\,{\rm eV}
(Klapdor-Kleingrothaus \etal 2002; although see also Aalseth \etal
2002; Feruglio, Strumia \& Vissani 2002) and the neutrino mass range
inferred from anthropic arguments by Tegmark \& Vilenkin (2003).

\begin{figure}
\vspace{0.5cm} \hbox{
\hspace{-0.0cm}\psfig{figure=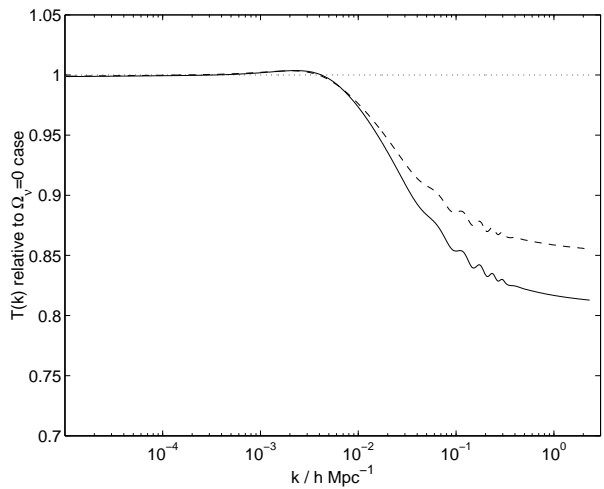,width=.45\textwidth,angle=0}}
\caption{The suppression of the matter power spectrum as a function of 
scale due to massive neutrinos, relative to the massless neutrino
case. The solid curve shows the suppression for
$\Omega_{\nu}h^2=0.0059$, the peak-probability value from the
CMB+2dF+$f_{\rm gas}$+XLF analysis, allowing for tensor components.
The dashed curve shows the results
for $\Omega_{\nu}h^2=0.0045$, the result from the CMB+2dF+$f_{\rm
gas}$ data. All other parameters were fixed to their best-fit values
(Table~\ref{table:marginal1}).}\label{fig:nupks8}
\end{figure}

As was discussed in Section 1, our result on the neutrino mass
density originates primarily from the fact that the XLF provides
a robust constraint on $\sigma_8$ for a given value of $\Omega_{\rm
m}$, while the CMB data predict $\sigma_8$ as a function
of the neutrino mass. The primary effect of the introduction
of massive neutrinos is to suppress the formation and growth of
structure on small scales ($k \approxlt 0.026(m_\nu/1\,{\rm
eV})^{0.5}\Omega_{\rm m}^{0.5}\,h$ Mpc$^{-1}$; Hu, Eisenstein \&
Tegmark 1998). The large momenta of such neutrinos prevents them
from clustering with the cold mass components and they  
stream freely. The suppression of the matter power spectrum 
as a function of scale in the case of massive neutrinos, 
relative to the massless neutrino case, is shown in  Fig.~\ref{fig:nupks8}.

Since the CMB data normalize the power spectrum primarily  on large
scales, the amplitude of fluctuations on (relatively  small)
$8\,h^{-1}$\,Mpc scales, $\sigma_8$, inferred from the CMB data will
always be smaller in the presence of massive neutrinos ($\sim 15$ per
cent smaller for $\Omega_{\nu}h^2=0.0059$; Fig~\ref{fig:nupks8}).  In
contrast, the value  of $\sigma_8$ inferred from the cluster XLF
analysis is relatively insensitive to massive neutrinos, since
clusters measure the power on scales similar to $8\,h^{-1}$Mpc. (The
mean length scale corresponding to the  virial masses of the clusters
in the XLF study of Allen \etal 2003 is $13.6\,h^{-1}$Mpc.)  For
$\Omega_{\nu}h^2=0.0059$, the value of $\sigma_8$ inferred  from the
XLF study is reduced by only $\sim 1.3$ per cent when accounting for
the effects of the neutrinos.

We note that on scales significantly smaller than those of clusters,
accounting for the presence of neutrinos at the measured level will
increase the inferred value of $\sigma_8$ by a small amount ($\sim
2.1$ per cent for measurements on $2\,h^{-1}$Mpc scales, for
$\Omega_{\nu}h^2=0.0059$). This may be relevant to studies of cosmic
shear and the Lyman$-\alpha$ forest.  Note, however, that  non-linear
corrections to the matter power spectrum also become important on such
scales (\eg Smith \etal 2003).

In what follows, we examine the most significant assumptions and
systematic uncertainties affecting our result on the neutrino mass.
These primarily concern the data sets used and the parameters varied
in the analysis.

\vspace{0.2cm}
\noindent$\bullet$
{\bf Reliability of XLF constraints.} As was mentioned above, the
result on $\sigma_8$ as a function of $\Omega_{\rm m}$ from the
cluster XLF study is crucial to our measurement of the neutrino mass.
Systematic uncertainties in the XLF analysis are discussed by Allen
\etal (2003). The main uncertainty in that work lies in the
mass-luminosity relation used to convert model mass functions into
X-ray luminosity functions, which are then compared with the
observations. Allen \etal (2003) find $\sigma_8=0.69\pm0.04$ for
$\Omega_{\rm m}=0.3$, accounting  for uncertainties in the
normalization, slope and scatter of the mass-luminosity
relation. Replacing the Allen \etal (2003) XLF constraints with  a
Gaussian probability density function (pdf) with
$\sigma_8=0.69\pm0.04$ leads to similar results on the neutrino mass.

We have examined the effects on the measured neutrino mass of
introducing additional, systematic offsets in the mass-luminosity
relation.  For example, replacing the Allen \etal (2003) XLF
constraints with a Gaussian pdf with $\sigma_8=0.77\pm0.04$ (which
corresponds to raising the normalization of the mass-luminosity curve
by a factor  $\sim 1.3$) reduces the neutrino mass inferred from the
CMB+2dF+$f_{\rm gas}$+(replaced) XLF data to $m_\nu
=0.14^{+0.10}_{-0.09}$\,eV. Conversely, lowering the normalization of
the mass-luminosity relation by a similar factor using a Gaussian pdf
with $\sigma_8=0.62\pm0.04$ increases the neutrino mass to $m_\nu
=0.26^{+0.08}_{-0.10}$\,eV. (Note that the inclusion of the
CMB+$f_{\rm gas}$ data maintains $\Omega_{\rm m}\sim0.3$.)

The results on the neutrino mass are not sensitive to moderate changes
in the uncertainty in $\sigma_8$ from the XLF data. Approximately
doubling the uncertainty in $\sigma_8$ using a Gaussian pdf with
$\sigma_8=0.69\pm0.08$ still gives $m_\nu
=0.19\pm0.10$\,eV with $\Omega_{\nu}h^2>0$ at more than $99$
per cent confidence for the CMB+2dF+$f_{\rm gas}$+(replaced) XLF data
(allowing for tensor components).  Only if the uncertainty in
$\sigma_8$ is quadrupled from its nominal value of 0.04 does the
detection of a non-zero neutrino mass drop below the 95 per cent
confidence level.

\vspace{0.2cm}
\noindent$\bullet$
{\bf Reliability of CMB polarization data.}  The CMB
temperature-polarization correlation power spectrum helps to constrain
the optical depth to reionization, $\tau$. The most probable values
for $\tau$ found from this analysis and by the WMAP team (Kogut  \etal
2003; Spergel \etal 2003) are higher than were  expected prior to the
release of the WMAP results. If we  were to exclude the CMB
polarization data, then the CMB  would constrain only the degenerate
combination  $\sigma_8 {\rm e}^{-\tau}$.  Since this product decreases with
increasing $\tau$, the CMB+2dF+$f_{\rm gas}$ constraint could also  be
matched with the XLF constraint on $\sigma_8$ if the value of $\tau$
were lowered. For example, if  we constrain $0.04<\tau<0.07$ in the
analysis of the  CMB+2dF+$f_{\rm gas}$ data, then we obtain
$\sigma_8=0.73\pm0.07$.
In this case, when combining with the XLF data, a negligible neutrino
mass falls on the $\sim 96$ instead of $99$ per cent confidence limit.

We note that the measured optical depth to reionization,
$\tau=0.09^{+0.06}_{-0.05}$ from our default analysis of  the full
CMB+2dF+$f_{\rm gas}$+XLF data set, is lower, and has larger
uncertainties, than the value quoted in the abstracts of the WMAP
papers ($0.17\pm0.04$; Kogut \etal 2003;  Spergel \etal 2003). Taken
together with the constraints on $n_{\rm S}$, our results may be 
slightly easier to explain within the context of standard UV reionization
models, even  allowing for the suppression of power on small scales
due to  the free-streaming action of massive neutrinos.

\vspace{0.2cm}
\noindent$\bullet$
{\bf Reliability of the 2dFGRS.} The 2dFGRS constraints used in this
paper rely on the assumption of linear bias. This issue was discussed
by Elgaroy \& Lahav (2003). However, the main point  here is that our
result on the neutrino mass is largely unchanged when removing the
2dFGRS constraint (Section~\ref{section:neutrinos}).

\vspace{0.2cm}
\noindent$\bullet$
{\bf The effect of tensors.} Including the presence of primordial
gravity waves in our analysis increases power on the largest
scales. This  is reflected by the fact that a  fit to the
CMB+2dF+$f_{\rm gas}$+XLF data in the presence of tensor  components
gives a slightly higher scalar spectral index, $n_{\rm
S}=0.98^{+0.04}_{-0.03}$,  than is obtained in the absence of tensors
$n_{\rm S}=0.95\pm0.02$.  The higher scalar spectral index in the presence of
tensors slightly increases the power on intermediate scales, relative
to the largest scales and, therefore, also  slightly increases the
neutrino mass relative to the no-tensors case. However, the detection
of a non-zero  neutrino mass in the absence of tensors remains
significant (Section~\ref{section:neutrinos}).

We note that the most probable scalar spectral index obtained  from
our analysis is consistent with standard inflation models. The ratio
of the tensor to scalar components is also consistent  with
single-field, slow-roll inflation ($R=-8n_{\rm T}$).

\vspace{0.2cm}
\noindent $\bullet$
{\bf The effect of a running spectral index.} There has been recent
speculation about a non-power law primordial power spectrum. Evidence
for a running spectral index was found by Spergel \etal (2003), when
the WMAP and other CMB data were combined with  2dFGRS and
Lyman-$\alpha$ forest constraints.  It is interesting to consider
whether our constraint on a significant neutrino mass could be negated
by allowing for a non-zero value of $n_{\rm run}$.  For the
CMB+2dF+$f_{\rm gas}$ data, we find an $n_{\rm run}$ value consistent
with that of Spergel et al. (2003), but with wider error bars
(assuming  a negligible neutrino mass). This is mainly due to the fact
that we do not  use the  Lyman-$\alpha$ forest data in our
analysis. However, we find that although the constraint on $\sigma_8$
is  widened when including $n_{\rm run}$ as a free parameter,  the
lower limit on $\sigma_8$ is hardly changed. This acts in an opposite
sense to allowing freedom in the neutrino mass and does not improve
the agreement between  CMB+2dF+$f_{\rm gas}$ and XLF constraints shown
in  Fig.~\ref{fig:nucont}.

\vspace{0.2cm}
\noindent $\bullet$
{\bf The effect of quintessence.} We can lower the value of $\sigma_8$
allowed by the CMB data if we increase the dark energy equation of
state parameter, $w$, above the cosmological  constant-like value of
$w=-1$ (see \eg Bridle \etal 2003b). Repeating the analysis of CMB
data, but replacing the free neutrino mass by a free equation of state
parameter, we obtain $w\sim-0.6$ for $\sigma_8\sim0.7$. However,  such
a solution is disfavoured by the combination of CMB+$f_{\rm gas}$
data, which prefer a value for $w \sim -1$. Our result on
quintessence is in line with supernovae studies, which also require $w
\sim -1$ for $\Omega_{\rm m}\sim0.3$ (Tonry \etal 2003).

In future work, we will examine in more detail the constraints that
the combination of CMB and X-ray cluster data can place on
the dark energy density and $w$.

\section*{Acknowledgements}

We thank G. Efstathiou for the suggestion to examine the cosmological
constraints that can be  obtained from the combination of CMB and
X-ray galaxy cluster data. We thank G. Efstathiou, O. Elgaroy,
A. Fabian,  A. Lewis and J. Ostriker for helpful discussions.  We are
grateful to A. Lewis for making the CosmoMC code publicly available
and V. Eke for communicating the results from his simulations.  The
CMB+2dF chains with  variable $n_{\rm run}$ are taken from Bridle
\etal (2003a).  We thank the members of the X-ray group at the
Institute of Astronomy for their generosity with computing power and
R. Johnstone for help with optimizing the codes. SWA  acknowledges the
support of the Royal Society.

\end{document}